\documentclass[11pt]{article}

\usepackage[noinfoline,cw]{imsartCW}

\RequirePackage[OT1]{fontenc}
\RequirePackage{amsthm,amsmath}
\RequirePackage{natbib}
\RequirePackage[colorlinks,citecolor=blue,urlcolor=blue]{hyperref}

\usepackage{array}
\usepackage{multirow}
\usepackage{amsmath}
\usepackage{amssymb}
\usepackage{graphicx}

\startlocaldefs
\numberwithin{equation}{section}
\theoremstyle{plain}

\endlocaldefs

\begin{document}

\begin{frontmatter}
\title{A Weighted U Statistic for Association Analyses Considering Genetic
Heterogeneity}
\runtitle{Heterogeneity Weighted U}

\begin{aug}
\author{\fnms{Changshuai} \snm{Wei}\thanksref{t1,m1}\ead[label=e1]{changshuai.wei@unthsc.edu}},
\author{\fnms{Robert C.} \snm{Elston}\thanksref{t2,m2}\ead[label=e2]{robert.elston@cwru.edu}}
\and
\author{\fnms{Qing} \snm{Lu}\thanksref{t3,m3}\ead[label=e3]{qlu@epi.msu.edu}}

\thankstext{t1}{Assistant Professor of Biostatistics, Department of
Biostatistics and Epidemiology, University of North Texas Health Science Center, (Email: \href{mailto:changshuai.wei@unthsc.edu}{changshuai.wei@unthsc.edu}) }
\thankstext{t2}{Professor of Biostatistics, Department of Epidemiology
and Biostatistics, Case Western Reserve University. (Email: \href{mailto:robert.elston@cwru.edu}{robert.elston@cwru.edu })}
\thankstext{t3}{Corresponding Author, Assosicate Professor of Biostatistics,
Department of Epidemiology and Biostatistics, Michigan State University.(Email:
\href{mailto:qlu@epi.msu.edu}{qlu@epi.msu.edu})}
\runauthor{C. Wei et al.}

\affiliation{University of North Texas Health Science Center \thanksmark{m1}, Case Western Reserve University \thanksmark{m2} and Michigan State University\thanksmark{m3}}


\end{aug}

\begin{abstract}
Converging evidence suggests that common complex diseases with the
same or similar clinical manifestations could have different underlying
genetic etiologies. While current research interests have shifted toward uncovering rare variants and structural variations predisposing to human diseases, the impact of heterogeneity in genetic studies of complex diseases has been largely overlooked. Most of the existing statistical methods assume the disease
under investigation has a homogeneous genetic effect and could, therefore,
have low power if the disease undergoes heterogeneous pathophysiological
and etiological processes. In this paper, we propose a heterogeneity
weighted U (HWU) method for association analyses considering genetic heterogeneity. HWU can be applied to various types of
phenotypes (e.g., binary and continuous) and is computationally efficient
for high-dimensional genetic data. Through simulations, we showed the advantage of HWU when the underlying genetic etiology
of a disease was heterogeneous, as well as the robustness of HWU against different model assumptions (e.g., phenotype distributions). Using HWU, we conducted a genome-wide
analysis of nicotine dependence from the Study of Addiction: Genetics
and Environments (SAGE) dataset. The genome-wide analysis of nearly
one million genetic markers took 7 hours, identifying heterogeneous
effects of two new genes (i.e., \textit{CYP3A5} and \textit{IKBKB}) on nicotine dependence.
\end{abstract}


\end{frontmatter}

\section{Introduction}

Benefiting from high-throughput technology and ever-decreasing genotyping
cost, large-scale genome-wide and sequencing studies have become commonplace
in biomedical research. From these large-scale studies, thousands
of genetic variants have been identified as associated with complex
human diseases, some with compelling biological plausibility for a
role in the disease pathophysiology and etiology. Despite such success,
for most complex diseases the identified genetic variants account
for only a small proportion of the heritability. While substantial
efforts have shifted toward finding rare variants, gene-gene/gene-environment
interactions, structural variations, and other genetic variants accounting
for the missing heritability [\cite{Eichler2010}], there is a considerable
lack of attention being paid to genetic heterogeneity in the analysis
of complex human diseases.We define genetic heterogeneity as a genetic
variant having different effects on individuals or on subgroups of
a population (e.g., gender and ethnic groups). For instance, the effect
size and the effect direction of the genetic variant can be different
according to the individuals' genetic background, personal/demographic
characteristics and/or the sub-phenotype groups they belong to.

Substantial evidence from a wide range of diseases suggests that complex
diseases are characterized by remarkable genetic heterogeneity [\cite{Thornton-Wells2004,McClellan2010,Galvan2010}]
. Despite the strong evidence of genetic heterogeneity in human disease
etiology, investigating genetic variants with heterogeneous effects
remains a great challenge, primarily because: i) the commonly used
study designs (e.g. the case-control design) may not be optimal for
studying heterogeneous effects; ii) there is a lack of prior knowledge
that can be used to infer the latent population structure (i.e., heterogeneous
subgroups in the population); iii) replication studies are more challenging
and need to be carefully designed; and iv) computationally efficient
and flexible statistical methods for high-dimensional data analysis,
taking into account genetic heterogeneity, have not been well developed.
Most of the existing methods assume that the disease under investigation
is a unified phenotype with homogeneous genetic causes. When genetic
heterogeneity is present, the current methods will likely yield attenuated
estimates for the effects of genetic variants, leading to low power
of the study.

To account for genetic heterogeneity in association analyses, we propose
a heterogeneity weighted U, referred to as HWU. Because the new method
is based on a weighted U statistic, it assumes no specific distribution
of phenotypes; it can be applied to both qualitative and quantitative
phenotypes with various types of distributions. Moreover, HWU is computationally
efficient and has been implemented in a C++ package for high-dimensional
data analyses (\href{https://www.msu.edu/~changs18/software.html\#HWU}{https://www.msu.edu/$\sim$changs18/software.html\#HWU}).

\section{Method}

\subsection{Motivation from a Gaussian random effect model}

To motivate the idea of the heterogeneity weighted U, we first introduce
a Gaussian random effect model to test genetic association when considering
genetic heterogeneity. Assume the following random effect model, 
\[
Y_{i}=\mu+g_{i}\beta_{i}+\varepsilon_{i},\varepsilon_{i}\sim N(0,\sigma^{2}),
\]
where $Y_{i}$ and $g_{i}$ represent the phenotype and the single-locus
genotype of individual $i$, respectively. $g_{i}$ can be coded as
0, 1, and 2 (i.e., the additive model), or 0 and 1 (e.g., the dominant/recessive
model); $\beta_{i}$ is normally distributed, $\beta_{i}\sim N(0,\sigma_{b}^{2})$,
and $\varepsilon_{i}$ is the iid random error. Let $\kappa_{i,j}$
represent the background similarity or the latent population structure
for individuals $i$ and $j$. We assume that the more similar two
individuals are, the more similar are their genetic effects, i.e.,
$cov(\beta_{i},\beta_{j})=\kappa_{i,j}\sigma_{b}^{2}$.

We define $\beta=(\beta_{1},\cdots,\beta_{n})^{T}$, $Y=(Y_{1},\cdots,Y_{n})^{T}$,
$\varepsilon=(\varepsilon_{1},\cdots,\varepsilon_{n})^{T}$, $G=\{diag(g_{1},\cdots,g_{n})\}{}_{n\times n}$
, and $K=\{\kappa_{i,j}\}_{n\times n}$. The model can then be written
as: $Y=\mu+G\beta+\varepsilon,\varepsilon\sim N(0,\sigma^{2}I),\beta\sim N(0,\sigma_{b}^{2}K)$.
We denote $\delta=G\beta$, and rewrite the model as: $Y=\mu+\delta+\varepsilon,\varepsilon\sim N(0,\sigma^{2}I),\delta\sim N(0,\sigma_{b}^{2}GKG)$.
A score test statistic can be formed to test the variance component
$\sigma_{b}^{2}=0$, 
\[
T=\tilde{Y}^{T}GKG\tilde{Y},
\]
where $\tilde{Y}_i=(Y_{i}-\mu)/\sigma$ is the standardized residual
under the null. We can partition the test statistic $T$ into two
parts, $T=\sum_{i\neq j}\kappa_{i,j}g_{i}g_{j}\tilde{Y}_{i}\tilde{Y}_{j}+\sum_{i=1}^{n}g_{i}^{2}\tilde{Y}_{i}^{2}$
, where the first summation is closely related to the weighted U statistic
introduced below.

\subsection{Heterogeneity weighted U}

The Gaussian random effect model assumes a normal distribution. In
order to consider phenotypes with various distributions and modes
of inheritance, we develop a heterogeneity weighted U with rank-based
U kernels and flexible weight functions. We first order the subjects
according to their phenotypic values $Y_{i}$ and assign subject scores
based on their ranks, denoted by $R_{i}$, $i=1,\cdots,n$. When there
are ties in the sample, we assign the averaged rank. For example,
in a case-control study with $N_{0}$ controls ($Y_{i}=0$) and $N_{1}$
cases ($Y_{i}=1$), all the controls are assigned a score $(N_{0}+1)/2$.
The phenotypic similarity between subjects $i$ and $j$ can be defined
as, 
\[
S_{i,j}=h(R_{i},R_{j}),
\]
where $h(\cdot,\cdot)$ is a two degree mean zero symmetric kernel
function (i.e., $h(R_{i},R_{j})=h(R_{j},R_{i})$ and $E_{F}(h(R_{i},R_{j}))=0$
) that satisfies the finite second moment condition, $E_{F}(h^{2}(R_{i},R_{j}))<\infty$,
and the degenerate kernel condition, $var(E(h(R_{i},R_{j})|R_{j}))=0$.
In this paper, we choose $h(R_{i},R_{j})=\sigma_{R}^{-2}(R_{i}-\mu_{R})(R_{j}-\mu_{R})$,
where $\mu_{R}=E(R)$ and $\sigma_{R}^{2}=var(R)$. Let $G_{i}=(g_{i,1},\cdots,g_{i,Q})$
denote the multiple genetic variants for individual $i$. We further
define a weight function to measure the genetic similarity under the
latent population structure $\kappa_{i,j}$, 
\[
w_{i,j}=\kappa_{i,j}f(G_{i},G_{j}),
\]
where $f(G_{i},G_{j})$ represents the genetic similarity calculated
based on the genetic variants of interest. We can then form the heterogeneity
weighted U, referred to as HWU, 
\[
U=2\sum_{1\leq i<j\leq n}w_{i,j}S_{i,j},
\]
to evaluate the association between
the phenotype and the genetic variants, considering the latent population
structure.

Thus HWU is a summation, over all pairs of individuals, of their phenotypic
similarities weighted by their genetic similarities. Under the null
hypothesis of no association, the phenotypic similarity is unrelated
to the genetic similarity. Because the phenotypic similarity has mean
0 (i.e., $E_{F}(R_{i},R_{j}) = 0$), the expectation of HWU is 0. Under
the alternative, the phenotypic similarities should increase as the
genetic similarities increases. The positive phenotypic similarities
are more heavily weighted and the negative phenotypic similarities
are more lightly weighted, leading to a positive value of HWU under
the alternative.

\subsection{Asymptotic distribution of heterogeneity weighted U}

To assess the significance of the association, a permutation test
can be used to calculate a p-value for HWU. However, for high-dimensional
data, the permutation test could be computationally intensive. Therefore,
we derive the asymptotic distribution of HWU under the null hypothesis.

The asymptotic properties of the un-weighted U statistic (i.e., $w_{i,j}\equiv1$)
are well established [\cite{Hoeffding1948,Serfling1981}]. When the kernel
is non-degenerate ( $var(E(h(R_{i},R_{j})|R_{j}))>0$ ), the limiting
distribution is normal. When the kernel is degenerate ( $var(E(h(R_{i},R_{j})|R_{j}))=0$
), the limiting distribution is a sum of independent chi-square variables.
However, the limiting distribution of the weighted U statistic depends
on both the weight function and the kernel function [\cite{ONeil1993}].
Because non-normality also occurs for a non-degenerate kernel with
certain weight functions, we use the degenerate kernel for HWU to
obtain a unified form of limiting distribution, as shown in the following
derivation.

We first expand the kernel function $h(\cdot,\cdot)$ as the sum of
products of its eigenfunctions. Let $\{\alpha_{t}\}$ and $\{\varphi_{t}(\cdot)\}$
denote the eigenvalues and the corresponding ortho-normal eigenfunctions
of the kernel. We can write $h(\cdot,\cdot)$ as $h(R_{i},R_{j})=\sum_{t=1}^{\infty}\alpha_{t}\varphi_{t}(R_{i})\varphi_s(R_{j})$
, and the weighted U as, 
\[
U=\sum_{i\neq j}w_{i,j}\sum_{t=1}^{\infty}\alpha_{t}\varphi_{t}(R_{i})\varphi_s(R_{j}),
\]
where 
\[
E(\varphi_{t}(R_{i})\varphi_s(R_{j}))=\begin{cases}
1, & t=s\text{ and }i=j\\
0, & \text{otherwise. }
\end{cases}
\]
By exchanging the two summations, ($\sum_{t=1}^{\infty}\alpha_{t}\sum_{i\neq j}w_{i,j}\varphi_{t}(R_{i})\varphi_s(R_{j})$),
the weighted U statistic is an infinite sum of quadratic forms and
can be approximated by a linear combination of chi-square random variables [\cite{Dewet1973,Shieh1994}].
Letting $W=\{w_{i,j}\}_{n\times n}$ be the weight matrix with all
diagonal element equal to 0, the limiting distribution can be written
as 
\[
U\sim\sum_{t=1}^{\infty}\alpha_{t}\sum_{s=1}^{n}\lambda_{s}(\chi_{1,ts}^{2}-1),
\]
where $\{\lambda_{s}\}$ are the eigenvalues of the weight matrix
and $\{\chi_{1,ts}^{2}\}$ are iid chi-square random variables with
1 df. In this paper, we use a cross product kernel, $h(R_{i},R_{j})=\sigma_{R}^{-2}(R_{i}-\mu_{R})(R_{j}-\mu_{R})$.
In this case, the expansion of $h(\cdot,\cdot)$ can be simplified
to $h(R_{i},R_{j})=\alpha_{1}\varphi_{1}(R_{i})\varphi_{1}(R_{j})$,
where $\alpha_{1}=1$ and $\varphi_{1}(R)=\sigma_{R}^{-1}(R-\mu_{R})$.
Using this representation and the fact that $\sum_{s=1}^{n}\lambda_{s}=0$,
the limiting distribution can be simplified to $U\sim\sum_{s=1}^{n}\lambda_{s}\chi_{1,s}^{2}$.
We also note that the parameters $\mu_{R}$ and $\sigma_{R}^{2}$
are unknown and need to be estimated from the data, which influences
the limiting distribution of HWU [\cite{Dewet1987,Shieh1997}]. Taking
the parameter estimation into account, the limiting distribution can
be expressed as a weighted sum of independent chi-squared variables,
$U\sim\sum_{s=1}^{n}\lambda_{1,s}\chi_{1,s}^{2}$ (Appendix \ref{app}), where $\{\lambda_{1,s}\}$ are the eigenvalues of the matrix $(I-J)W(I-J)$
, in which $I$ is an identity matrix and $J$ is a matrix with all
elements equal to $1/n$.

The HWU described above can also be modified to allow for covariate
adjustment. Suppose $Z_{n\times p}=(1,z_{1},\cdots,z_{p})$ is the
covariate matrix. In the cross product kernel of HWU, we can calculate
the estimators of $\mu_{R}$ and $\sigma_{R}^{2}$ as $\hat{\mu}_{R}=PR$
and $\hat{\sigma}_{R}^{2}=(R-\hat{\mu}_{R})^{T}(R-\hat{\mu}_{R})/(n-p-1)$,
where $P=Z(Z^{T}Z)^{-1}Z^{T}$. The limiting distribution can then
be written as $U\sim\sum_{s=1}^{n}\lambda_{1,s}^{*}\chi_{1,s}^{2}$
, where $\{\lambda_{1,s}^{*}\}$ are the eigenvalues of the matrix
$(I-P)W(1-P)$.

Davies\textquoteright{} method [\cite{Davies1980}] can be used to calculate
the p-value for the association test. When the calculation involves
large matrix eigen-decomposition, we use the state-of-the-art algorithm
nu-TRLan [\cite{Wu2000}] to improve the computational efficiency.

\subsection{Weighting schemes}

The weight function comprises two components, $\kappa_{i,j}$ and
$f(G_{i},G_{j})$ . $\kappa_{i,j}$ measures the latent population
structure, which could be inferred from related covariates. Depending
on the type of data, different functions can be used to calculate
$\kappa_{i,j}$. For instance, we can apply the genome-wide averaged
IBS function on GWAS data and the genome-wide weighted average IBS
(WIBS) function on sequencing data to calculate $\kappa_{i,j}$ [\cite{Astle2009}].
For environmental covariates, we can calculate $\kappa_{i,j}$ based
on Euclidian distance [\cite{Jiang2011}]. Given environmental covariates,
we first standardize each covariate according to its mean and standard
deviation, denoted by $x_{d}$ ($x_{d}=(x_{d,1},\cdots,x_{d,n})^{T}$
, $d=1,2,\cdots,D$ ), and then calculate $\kappa_{i,j}=exp(-(x_{i}-x_{j})R(x_{i}-x_{j})^{T})$,
where $R$ is used to reflect the relative importance (e.g., $R=\{diag(\omega_{d})\}_{D\times D}$
in which $\omega_{d}$ measures the importance) or inner correlation
(e.g., $R=(\frac{1}{n}\sum_{i=1}^{n}x_{i}^{T}x_{i})^{-1}$) of the
covariates. 

$f(G_{i},G_{j})$ measures the genetic similarity. For a single-locus
model, we can use the cross product $f(G_{i},G_{j})=f(g_{i},g_{j})=g_{i}g_{j}$
when the effect is additive. Otherwise, we can use $f(G_{i},G_{j})=f(g_{i},g_{j})=1(g_{i}=g_{j})$
for an unspecified mode of inheritance, where $1(\cdot)$ is the indicator
function. The above measurements can be easily extended to handle
$Q$ multiple markers by using $f(G_{i},G_{j})=\sum_{q=1}^{Q}g_{q,i}g_{q,j}$.

The weight function $w_{i,j}$ can also be specified for different
purposes. For instance, if we choose $w_{i,j}=f(G_{i},G_{j})$ (i.e.,
$\kappa_{i,j}\equiv1$), then the weighted U tests the association
without consideration of genetic heterogeneity. We refer to this statistic
as the non-heterogeneity weighted U (NHWU). Furthermore, we can construct
a statistic to test the presence of the heterogeneity effect, referred
to as the pure-heterogeneity weighted U (PHWU), by setting $w_{i,j}^{*}=(\kappa_{i,j}-\bar{\kappa})f(G_{i},G_{j})$,
where $\bar{\kappa}=\frac{1}{n^{2}}\sum_{i=1}^{n}\sum_{j=1}^{n}\kappa_{i,j}$.

\section{Result}

\subsection{Simulations}

In simulation I and simulation II, we simulated various cases of genetic
heterogeneity and compared the proposed HWU test with two other tests,
NHWU and the likelihood ratio test using the conventional generalized
linear model (GLM). In simulation III, we investigated the robustness
of HWU to non-normal distributions and mis-specified weight functions.
In all sets of simulations, unless otherwise specified we used Euclidian-distance-based
$\kappa_{i,j}$ by setting $R=I$ and cross-product-based $f(g_{i},g_{j})$
to form the weight function. For each simulation setting, we simulated
1000 replicate datasets, each having a sample size of 1000. Power
and type 1 error of the methods were calculated based on the proportion
of p-values in the 1000 replicates smaller than or equal to 0.05.

\subsubsection{Simulation I}

In this simulation, we assumed two sub-populations, and considered
both continuous and binary phenotypes. We simulated binary phenotypes
using the logistic model,

\[
logit(P(y_{i,j(i)}=1))=\mu+g_{i,j(i)}\beta_{i},
\]
where $i$ and $j(i)$ represented respectively the $i$-th
subpopulation and $j$-th individual in the $i$-th sub-population.
Additionally, we introduced a covariate $x_{i,j(i)}=a_{i}+\delta_{i,j(i)}$,
$\delta_{i,j(i)}\sim N(0,\sigma_{c}^{2})$, from which we infer the
latent sub-populations. Continuous phenotypes were simulated similarly
by using a linear regression model. The value of the regression coefficient
$\beta_{i}$ for different models was listed in Table 1, while the
details of the simulation were described in Supplementary Appendix
A.

No substantial inflation of type I error was detected for any of the
three methods (Table \ref{Table1}). In the presence of genetic heterogeneity
(i.e., T1, T3, and T4 in Table \ref{Table1}), HWU outperformed NHWU
and GLM, especially when the genetic effects for the two sub-populations
were in the opposite direction (i.e., T1). In such a case, NHWU and
GLM could barely detect any genetic effect, while HWU had high statistical
power to detect the association. In the absence of genetic heterogeneity
(i.e., T2 in Table \ref{Table1}), HWU remained comparable in performance
to NHWU and GLM. We also noted that, in the absence of genetic heterogeneity
(i.e., T2), the non-parametric NHWU had almost identical power to
GLM.

\begin{table}

\caption{Type I error and power comparison of three methods when there are
two heterogeneous sub-populations}
\label{Table1}
\begin{center}

\begin{tabular}{ccccccccccc}
\hline 
\multirow{3}{*}{Model\textsuperscript{1}} & \multicolumn{5}{c}{Binary Phenotype} & \multicolumn{5}{c}{Continuous Phenotype}\tabularnewline
\cline{2-11} 
 & \multicolumn{2}{c}{Effect\textsuperscript{2}} & \multicolumn{3}{c}{Type I error/Power} & \multicolumn{2}{c}{Effect} & \multicolumn{3}{c}{Type I error/Power}\tabularnewline
\cline{2-11} 
 & $\beta_{1}$ & $\beta_{2}$ & HWU & NHWU & GLM & $\beta_{1}$ & $\beta_{2}$ & HWU & NHWU & GLM\tabularnewline
\hline 
\hline 
Null & 0 & 0 & 0.046 & 0.051 & 0.051 & 0 & 0 & 0.052 & 0.049 & 0.049\tabularnewline
\hline 
T1 & -0.1 & 0.1 & 0.085 & 0.07 & 0.07 & -0.1 & 0.1 & 0.241 & 0.062 & 0.07\tabularnewline
 & -0.3 & 0.3 & 0.455 & 0.069 & 0.068 & -0.3 & 0.3 & 0.972 & 0.157 & 0.173\tabularnewline
 & -0.5 & 0.5 & 0.899 & 0.084 & 0.084 & -0.5 & 0.5 & 1 & 0.376 & 0.417\tabularnewline
\hline 
T2 & 0.1 & 0.1 & 0.122 & 0.149 & 0.15 & 0.1 & 0.1 & 0.284 & 0.368 & 0.384\tabularnewline
 & 0.3 & 0.3 & 0.601 & 0.743 & 0.75 & 0.3 & 0.3 & 0.999 & 1 & 1\tabularnewline
 & 0.5 & 0.5 & 0.978 & 0.993 & 0.993 & 0.5 & 0.5 & 1 & 1 & 1\tabularnewline
\hline 
T3 & 0 & 0.2 & 0.107 & 0.102 & 0.102 & 0 & 0.2 & 0.32 & 0.273 & 0.282\tabularnewline
 & 0 & 0.4 & 0.378 & 0.307 & 0.31 & 0 & 0.4 & 0.902 & 0.767 & 0.803\tabularnewline
 & 0 & 0.6 & 0.718 & 0.582 & 0.586 & 0 & 0.6 & 0.999 & 0.978 & 0.984\tabularnewline
\hline 
T4 & -0.1 & 0.3 & 0.239 & 0.091 & 0.092 & -0.1 & 0.3 & 0.718 & 0.161 & 0.18\tabularnewline
 & 0.1 & 0.3 & 0.304 & 0.377 & 0.381 & 0.1 & 0.3 & 0.822 & 0.89 & 0.906\tabularnewline
 & -0.3 & 0.5 & 0.72 & 0.069 & 0.069 & -0.3 & 0.5 & 0.997 & 0.059 & 0.076\tabularnewline
\hline 
\end{tabular}

\end{center}

\textsuperscript{1}Various scenarios of heterogeneity were considered
in the simulation, including no genetic effect for both sub-populations
(Null), the same effect size but with different directions (T1), the
same effect size with the same direction (T2), no genetic effect for
one sub-population but having a genetic effect for the other (T3),
and different effect sizes with the same or different directions (T4).

\textsuperscript{2}Single-locus effects for the 2 sub-populations,
where the effect for the sub-population 1 denoted by $\beta_{1}$
and the effect for the sub-population 2 denoted by $\beta_{2}$. 
\end{table}

We also investigated the performance of the three methods when the
underlying phenotype distribution and the modes of inheritance were
unknown (Supplementary Simulation I). Overall, HWU outperformed the
other two methods. In particular, when the phenotype was non-normal,
both HWU and NHWU had higher power than GLM (Supplementary Table S1).
By using $f(g_{i},g_{j})=1(g_{i}=g_{j})$, HWU was robust to the disease
model when the mode of inheritance was unknown, e.g., heterozygote
effect (Supplementary Table S2 and S3).

\begin{figure}
\caption{Power comparison of three methods for a binary phenotype with 20 heterogeneous
sub-populations}
\label{Fig1}

\begin{center}
 \includegraphics[width=1.0\linewidth,height=0.5\linewidth]{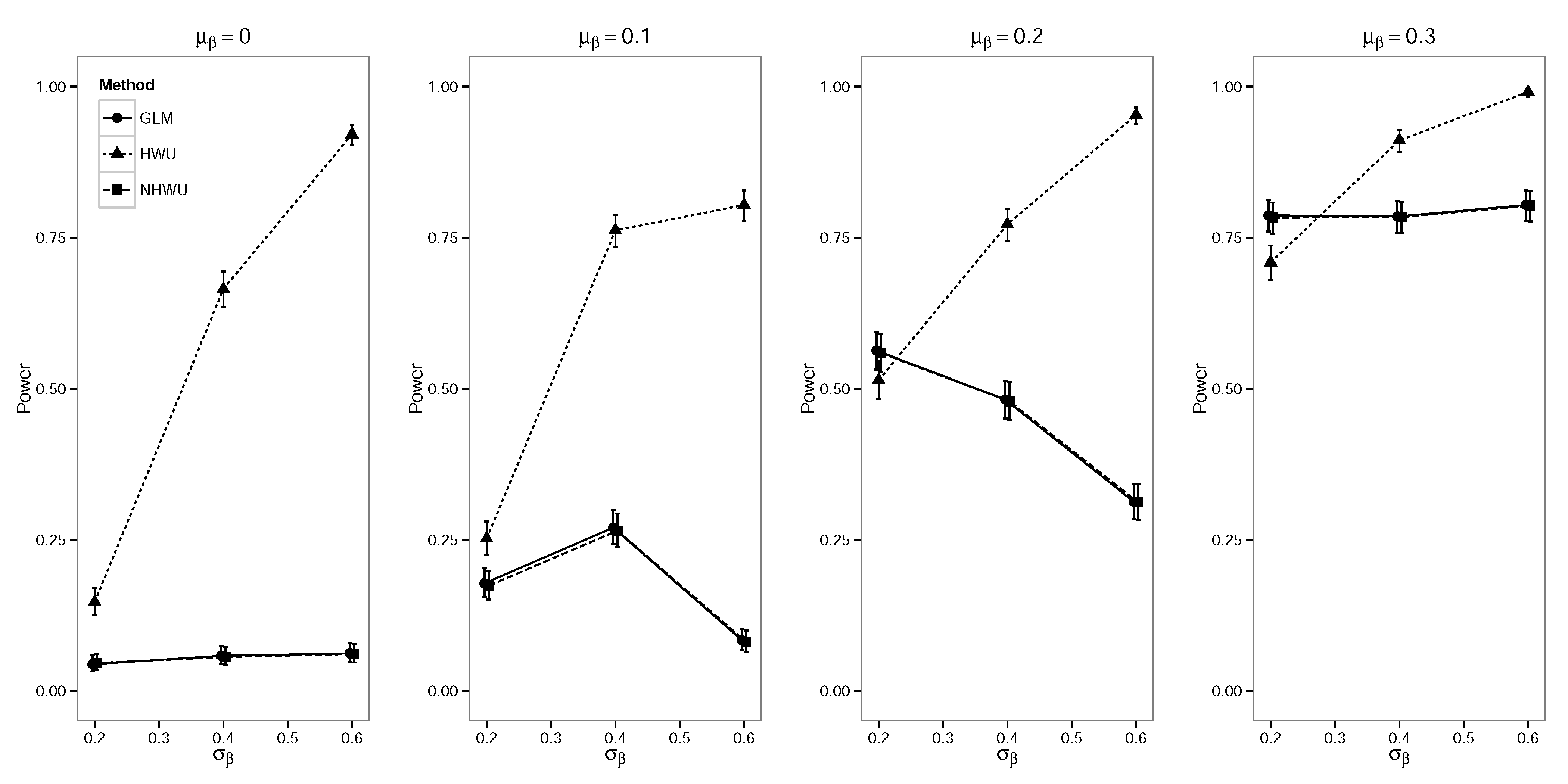}
\end{center}
{*}the genetic effect, $\beta_{i}$, for the i-th sub-population was
sampled from a uniform distribution with mean $\mu_{\beta}$ and variance
$\sigma_{\beta}^{2}$.
\end{figure}

\begin{figure}
\caption{Power comparison of three methods for a continuous phenotype with
20 heterogeneous sub-populations }
\label{Fig2}
\begin{center}
\includegraphics[width=1.0\linewidth,height=0.5\linewidth]{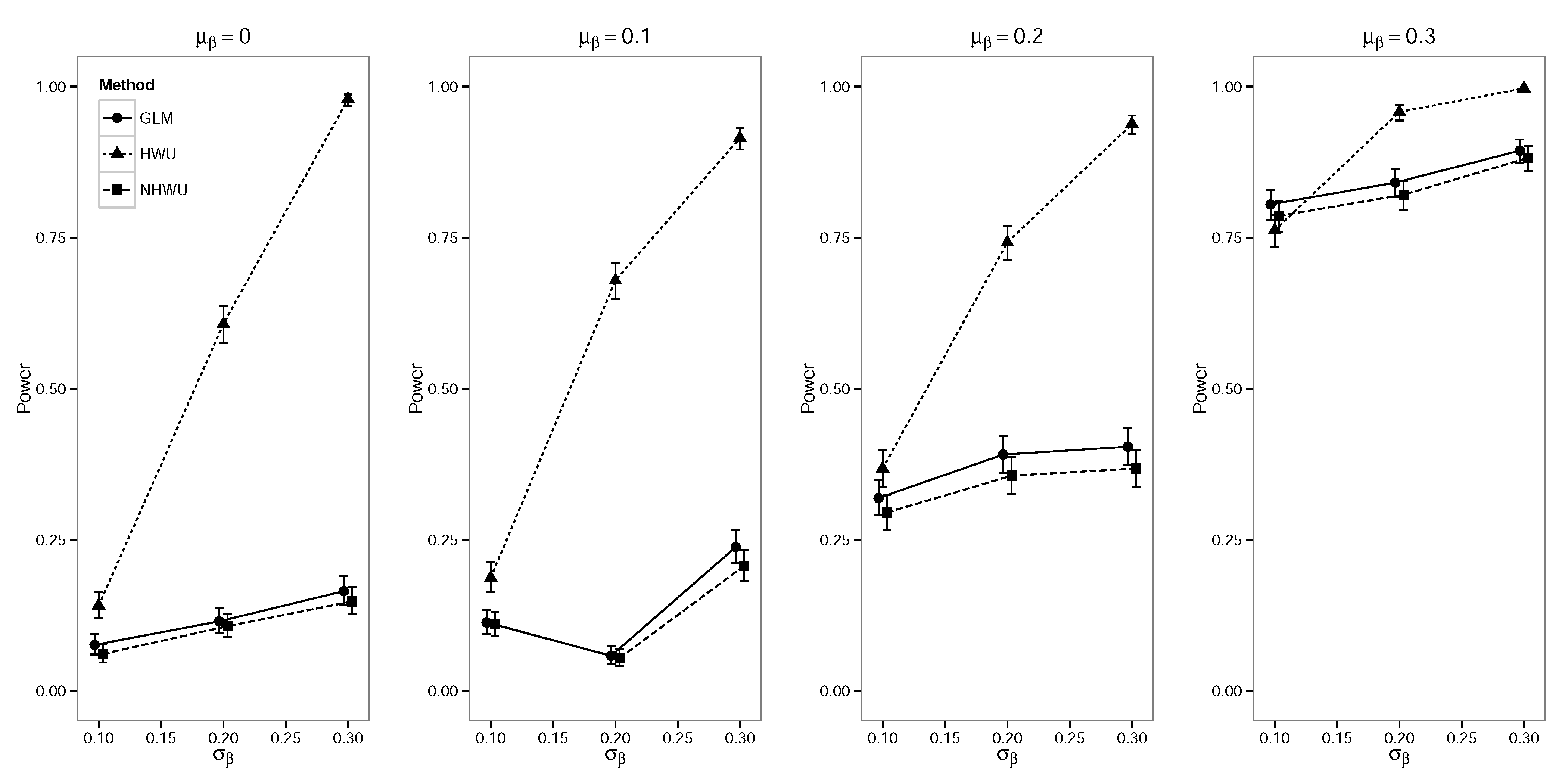} 
\end{center}
{*}the genetic effect, $\beta_{i}$, for the i-th sub-population was
sampled from a uniform distribution with mean $\mu_{\beta}$ and variance
$\sigma_{\beta}^{2}$.
\end{figure}

\subsubsection{Simulation II}

In simulation II, we used the same simulation model as in simulation
I, but considered a more complicated latent population structure by
increasing the number of sub-populations to 20, and sampling $\beta_{i}$
( $i=1,2,\cdots,20$ ) from a uniform distribution with mean $\mu_{\beta}$
and variance $\sigma_{\beta}^{2}$. We simulated 25 covariates, $x_{i,j(i)}^{d}=a_{i}^{d}+\delta_{i,j(i)}$,
$\delta_{i,j(i)}\sim N(0,\sigma_{c}^{2})$, ( $d=1,2,\cdots,25$ ),
to generate the latent population structure (Supplementary Appendix
B). No substantial inflation of type I error was detected for any
of the three methods at the 0.05 level (Supplementary Table S4). Through
simulation, we demonstrated that HWU outperformed NHWU and GLM for
both binary (Figure \ref{Fig1}) and continuous (Figure \ref{Fig2})
phenotypes. In the presence of genetic heterogeneity (i.e.,when $\sigma_{\beta}/\mu_{\beta}$
is large), HWU attained higher power than NHWU and GLM. When the genetic
heterogeneity was negligible (i.e., when $\sigma_{\beta}/\mu_{\beta}$
is small), HWU had comparable performance to NHWU and GLM. When the
average genetic effect ( $\mu_{\beta}$ ) increased, all three methods
gained power. Nevertheless, when the variance of the genetic effect
( $\sigma_{\beta}$ ) increased, only HWU gained substantial increase
in power. We also investigated the performance of HWU when the covariates
could not accurately infer the latent population structure. For such
purpose, we investigated the power of HWU as the noise parameter $\sigma_{c}^{2}$
changed. The result showed that the power of HWU decreased as the
\textquotedblleft{}noise\textquotedblright{} increased
(Supplementary Table S5).

In practice, the nature of the latent population structure may not
be \textquotedblleft{}categorical\textquotedblright{}. Therefore,
we also simulated genetic effects using a random effect model, where
effects were different for each subject (Supplementary Simulation
II). The three methods had comparable power when the genetic heterogeneity
was negligible. Nevertheless, as the genetic heterogeneity increased,
there was a clear advantage of HWU over NHWU and GLM (Supplementary
Figures S1 and S2).

\subsubsection{Simulation III}

In simulation III, we first investigated the robustness of HWU against
different non-normal phenotype distributions. In order to separate the influence of heterogeneity and phenotype distribution, we compared HWU with its \textquotedblleft{}parametric alternative\textquotedblright{},
the variance component score test (VCscore), instead of GLM. We simulated the phenotype
using a random effect model, 

\[
y_{i}=\mu+Z_{i}\alpha+g_{i}\beta_{i}+\varepsilon_{i},\varepsilon_{i}\sim F,
\]
where $Z_{i}$ denotes covariates for subject $i$, $\alpha$ denotes
covariate effects and $F$ followed a non-normal distribution (Supplementary Appendix C). We
simulated three types of non-normal distribution for $F$, 1) t ditributions
with $df=2$, 2) Cauchy distribution, and 3) a mixture of normal and
chi-squared distribution. For each distribution, we simulated model
with confounding effects and without confounding effect, where confounding
effect is simulated by generating $Z_{i}$ that is correlated with $g_{i}$. Meanwhile, $Z_{i}$ is also correlated with $y_i$ since $\alpha \neq 0$.
We included $Z$ in the analysis for both HWU and VCscore, and summarize
the Type I errors in Table \ref{Table2}. No substantial inflation
of type I error was detected for HWU for 3 non-normal distributions,
regardless of whether there were confounding effects. VCscore is robust
against mixture of normal and chi-squared distribution, but have inflated
type I error for heavy tailed distribution (e.g., Cauchy distribution).
If we did not include $Z$ in the analysis, both methods showed inflated
type I error when there were confounding effects(Supplementary Table
S6). Further investigations on power performance showed slightly more
advantage of HWU over VCscore for non-normal distributions (Supplementary
Table S7). 

\begin{table}
\caption{Type I error comparisons of HWU and VCscore under non-normal distributions}
\label{Table2}
\begin{center}
\begin{tabular}{ccccc}
\hline 
\multirow{2}{*}{Confounding Effect} & \multirow{2}{*}{Model} & \multicolumn{3}{c}{Distribution}\tabularnewline
\cline{3-5} 
 &  & Mixture & $t_{df=2}$ & Cauchy\tabularnewline
\hline 
\hline 
\multirow{2}{*}{No} & HWU & 0.041 & 0.057 & 0.057\tabularnewline
\cline{2-5} 
 & VCscore & 0.039 & 0.070 & 0.095\tabularnewline
\hline 
\multirow{2}{*}{Yes} & HWU & 0.053 & 0.060 & 0.062\tabularnewline
\cline{2-5} 
 & VCscore & 0.059 & 0.146 & 0.250\tabularnewline
\hline 
\end{tabular}
\end{center}
{*}the mixture distribution follows $a\chi_{df=1}^{2}+(1-a)N(5,1)$,
where $a\sim Bernoulli(0.6)$.
\end{table}

\begin{table}

\caption{Performance of HWU with a mis-specified weight function}
\label{Table2_1}

\begin{tabular}{cccccc}
\hline 
\multirow{2}{*}{Component} & \multicolumn{2}{c}{Mis-specification\textsuperscript{1}} & \multirow{2}{*}{Model\textsuperscript{2}} & \multicolumn{2}{c}{Method\textsuperscript{3}}\tabularnewline
\cline{2-3} \cline{5-6} 
 & Mis & True &  & HWU(mis) & HWU(true)\tabularnewline
\hline 
\hline 
\multirow{4}{*}{$f(g_{i},g_{j})$ } & \multirow{2}{*}{$\mathfrak{D}(g_{i},g_{j})$ } & \multirow{2}{*}{$g_{i}g_{j}$} & Null & 0.047 & 0.049\tabularnewline
 &  &  & Alt & 0.459 & 0.481\tabularnewline
\cline{2-6} 
 & \multirow{2}{*}{$1(g_{i}=g_{j})$ } & \multirow{2}{*}{$g_{i}g_{j}$} & Null & 0.048 & 0.052\tabularnewline
 &  &  & Alt & 0.451 & 0.51\tabularnewline
\hline 
\multirow{4}{*}{$\kappa_{i,j}$} & \multirow{2}{*}{$\frac{1}{D}x_{i}x_{j}^{T}$} & \multirow{2}{*}{$\mathfrak{D}(x_{i},x_{j})$} & Null & 0.05 & 0.058\tabularnewline
 &  &  & Alt & 0.174 & 0.505\tabularnewline
\cline{2-6} 
 & \multirow{2}{*}{$\mathfrak{D}(x_{i},x_{j})$} & \multirow{2}{*}{$\frac{1}{D}x_{i}x_{j}^{T}$} & Null & 0.046 & 0.053\tabularnewline
 &  &  & Alt & 0.072 & 0.515\tabularnewline
\hline 
\end{tabular}

\textsuperscript{1}\textquotedblleft{}Mis\textquotedblright{} represents
the misspeficied $f(g_{i},g_{j})$ or $\kappa_{i,j}$ when analyzing
simulated data, while \textquotedblleft{}True\textquotedblright{}
represent the true $f(g_{i},g_{j})$ or $\kappa_{i,j}$ in the corresponding
simulation setting. Here, $\mathfrak{D}(\cdot,\cdot)$ represents
the euclidian distance based weight, i.e., $\mathfrak{D}(g_{i},g_{j})=exp(-(g_{i}-g_{j})^{2})$
and $\mathfrak{D}(x_{i},x_{j})=exp(-\frac{1}{D}\sum_{d=1}^{D}(x_{d,i}-x_{d,j})^{2})$.

\textsuperscript{2}The error distribution was set as $t$ distribution with $df=2$. \textquotedblleft{}Null\textquotedblright{}
represents the null model with $\mu_{\beta}=0$ and $\sigma_{\beta}^{2}=0$;
\textquotedblleft{}Alt\textquotedblright{} represents the heterogeneous
effect model with $\mu_{\beta}=0$ and $\sigma_{\beta}^{2}=0.5$.

\textsuperscript{3}HWU(mis) represents the HWU model with a mis-specified
weight function, while HWU(true) represents the HWU model with the true
weight function.
\end{table}

We also investigated the performance of HWU when the weight function
was mis-specified (Table \ref{Table2_1}). In this simulation,
we considered 4 different scenarios, either with mis-specified $f(g_{i},g_{j})$
or mis-specified $\kappa_{i,j}$. Type I error rates were well controlled
when the weight function was mis-specified. However, we found the
power of HWU with a mis-specified weight function was lower than that
with a correct weight function, especially when $\kappa_{i,j}$ was
mis-specified (Table \ref{Table2_1}).

\subsection{Genome-wide association analysis of Nicotine Dependence}

We applied our methods to the Genome-wide association study (GWAS)
dataset from the Study of Addiction: Genetics and Environments (SAGE).
The SAGE is one of the largest and most comprehensive case-control
studies conducted to date aimed at discovering new genetic variants
contributing to addiction. We analyzed the number of cigarettes smoked
per day, categorized into 4 classes (0 for less than 10 cigarettes,
1 for 11-to-20 cigarettes, 2 for 21-to-30 cigarettes, and 3 for more
than 31 cigarettes). Prior to the statistical analysis, we reassessed
the quality of the genotype data. After undertaking a careful quality
control process (i.e., removing samples with missing phenotype data
and low-quality genetic markers), 2845 subjects and 949,658 single-nucleotide
polymorphisms (SNPs) remained for the analysis. The SAGE comprises
samples from both Caucasian and African-American populations. To make
the association analysis robust against confounding effects, we adjusted
for the first 20 principal components from the available genome-wide
genetic markers, as well as gender and race, in the analysis. 

\begin{table}

\caption{Top 10 nicotine dependence associated SNPs from the GWAS analysis
considering gender heterogeneity }
\label{Table3}
\begin{center}
\begin{tabular}{cccccc}
\hline 
\multirow{2}{*}{Name} & \multirow{2}{*}{Chr} & \multirow{2}{*}{Position} & \multirow{2}{*}{Gene} & \multicolumn{2}{c}{p-value}\tabularnewline
\cline{5-6} 
 &  &  &  & HWU & NHWU\tabularnewline
\hline 
\hline 
rs17078660 & 3 & 46160432 & NA, near \textit{FLT1P1} & $9.98\times10^{-9}$ & 0.017\tabularnewline
MitoA15302G & 26 & 15302 & NA & $1.69\times10^{-8}$ & 0.688\tabularnewline
rs7753843 & 6 & 67055504 & NA & $1.86\times10^{-8}$ & 0.571\tabularnewline
rs10493279 & 1 & 60368804 & NA, near \textit{C1orf87} & $1.88\times10^{-8}$ & 0.014\tabularnewline
rs4560769 & 8 & 42259961 & \textit{IKBKB} & $1.93\times10^{-8}$ & 0.062\tabularnewline
rs9694958 & 8 & 42275203 & \textit{IKBKB} & $2.82\times10^{-8}$ & 0.122\tabularnewline
rs776746 & 7 & 99108475 & \textit{CYP3A5} & $2.91\times10^{-8}$ & 0.241\tabularnewline
rs4646437 & 7 & 99203019 & \textit{CYP3A4} & $4.03\times10^{-8}$ & 0.512\tabularnewline
rs9694574 & 8 & 42279609 & \textit{IKBKB} & $4.63\times10^{-8}$ & 0.144\tabularnewline
rs4646457 & 7 & 99083016 & \textit{ZSCAN25} & $4.74\times10^{-8}$ & 0.138\tabularnewline
\hline 
\end{tabular}
\end{center}
\end{table}

Considering that the etiology of nicotine dependence has been shown
to be heterogeneous for gender [\cite{Li2003}], we used gender to infer
the latent population structure and assumed an additive effect to
compute . Using HWU, the genome-wide scanning of 949,658 SNPs on the
SAGE dataset was completed in about 7 hours by parallel computation
on 19 cores. The top 10 SNPs having the strongest association with
nicotine dependence are listed in Table \ref{Table3}. Among the 10
SNPs, 3 SNPs (i.e., rs4560769, rs9694574, and rs9694958) are located
within the gene \textit{IKBKB}, while another 3 SNPs (i.e., rs4646437, rs4646457,
rs776746) are located within or near the gene \textit{CYP3A5}. The 3 SNPs related
to gene \textit{IKBKB} are in high linkage disequilibrium (LD), with the estimated
correlation ranging from 0.736 to 0.853. The highest association signal
was from rs4560769 (p-value$=1.93\times10^{-8}$). The 3 SNPs related
to gene \textit{CYP3A5} were also in high LD (correlation from 0.781 to 0.913),
among which rs4646437 had the strongest association with nicotine
dependence (p-value$=2.91\times10^{-8}$). To evaluate the sensitivity
of the results, we performed association tests using other weight
functions (Table \ref{Table3}). Using a homogeneity weight $w_{i,j}=g_ig_j$ (NHWU), none
of the 10 SNPs had a p-value smaller than 0.01. The difference between
HWU and NHWU indicated heterogeneous effects of the two genes on nicotine
dependence in males and females. Additional stratified analysis by
analyzing males and females separately also suggested this heterogeneous
effect of the two genes in males and females (Supplementary Real Data
Analysis). In addition to gender, we also investigated potential genetic
heterogeneity due to different ethnic and genetic backgrounds. In
these analyses, we considered the same covariates as those used in
the gender heterogeneity analysis. However, the results suggested
there was no strong evidence of genetic heterogeneity due to different
ethnic and genetic backgrounds (Supplementary Real Data Analysis).

\section{Discussion}

In recent years, U-statistic based methods have been gaining popularity
in genetic association studies due to their robustness and flexibility
[\cite{Schaid2005,Zhang2010}]. Yet, few methods have been developed
to model genetic heterogeneity, especially under the weighted U framework.
In this paper, we have proposed a flexible and computationally efficient
method, HWU, for high-dimensional genetic association analyses allowing
for genetic heterogeneity. With HWU, we were able to integrate the
latent population structure (inferred from genetic background or environmental
covariates) into a weight function and test heterogeneous effects
without stratifying the sample. Simulation studies were conducted
to compare the power of the proposed HWU method with methods that
do not model genetic heterogeneity (i.e., NHWU and GLM). In the presence
of genetic heterogeneity, HWU attained higher power than NHWU and
GLM. In the absence of genetic heterogeneity, HWU still had comparable
performance to NHWU and GLM. Unlike conventional methods, such as
GLM, our method was developed based on a nonparametric U statistic,
and therefore offers robust performance when the underlying phenotype
distribution and mode of inheritance are unknown.

In HWU, we use genome profiles or environmental covariates to build
the background similarity (i.e., the latent population structure $\kappa_{i,j}$
) and combine it with the genetic similarity to form the weight function
$f(G_{i},G_{j})$. We then evaluate its relationship with a phenotype
by using a weighted U statistic. Our method is different from testing
an interaction effect. The key difference is that, for HWU, we assume
there is a latent population structure that acts in some joint fashion
with the genetic variants, while in the usual interaction effect model
the genetic variants are assumed to interact with known variables.
Furthermore, our test has fewer degrees of freedom than usual interaction
tests. HWU is based on the idea that the more similar two subjects
are, the more similar are their genetic effects. The idea of relating
phenotype similarity to genotype similarity is not new. For example,
Tzeng et.al proposed a gene-trait similarity regression for multi-locus
association analysis [\cite{Tzeng2007}]. However, their method is based
on the usual regression framework and does not consider genetic heterogeneity. 

In this paper, we focus on a single-locus test with consideration
of genetic heterogeneity and assume an additive model. By modifying
the weight function, HWU can easily be extended to model a multi-locus
effect and other modes of inheritance (e.g., dominant/recessive effects).
The weight function also offers flexibility for constructing latent
population structure. Various similarity-based or distance-based functions
can be applied to informative environmental and genetic covariates
to infer the latent population structure. Although type I error is
generally controlled for a variety of weight functions, the choice
of an appropriate function to construct the latent population structure
could impact the power of HWU. In this article, we suggest a Euclidian-distance
based function, $\kappa_{i,j}=exp(-(x_{i}-x_{j})R(x_{i}-x_{j})^{T})$,
in which prior knowledge can be incorporated for potential power improvement.
Nevertheless, a cross product kernel (i.e., $\kappa_{i,j}=\frac{1}{D}x_{i}x_{j}^{T}$)
can also be used if the underlying model favors linearity. In the
scenario where multiple functions might be used to construct the latent
population structure, the optimal function could be chosen by using
a similar approach to that proposed by \cite{Lee2012}. 

Another advantage of our method is its computational efficiency. For
the analysis of high-dimensional data, we derived the asymptotic distribution
of the weighted U statistic and optimized the computational algorithm
(e.g. using efficient eigen-decomposition). The genome-wide analysis
of 949,658 SNPs took 7 hours and identified two genes, \textit{IKBKB} and \textit{CYP3A5}.
Although our analysis suggests that these two genes are associated
with nicotine dependence and have heterogeneous effects according
to gender, further study and biological experiments are needed to
confirm the association and to further investigate the potential function
of these two genes in nicotine dependence.

\appendix

\section{}

\subsection{Asymtotic distribution of HWU with parameter estimation}\label{app}
As showed in the main text, the limiting distribution of the weighted
U with a cross product kernel can be simplified to $U\sim\sum_{s=1}^{n}\lambda_{s}\chi_{1,s}^{2}$.
Taking the parameter estimation into account {[}\cite{Dewet1987,Shieh1997}{]}
, the limiting distribution becomes:

\[
U\sim\sum_{s=1}^{n}\lambda_{s}(\phi_{s}+c_{s}\phi_{0})^{2},
\]
where $\{\lambda_{s}\}$ are the eigenvalues from the eigen-decomposition
of $W=B\Lambda B^{T}$, in which $\Lambda=\{diag(\lambda_{s})\}_{n\times n}$
and $B=\{b_{i,j}\}_{n\times n}$. $\{\phi_{s}\}$ are i.i.d. standard
normal random variables. $\phi_{0}$ is also standard normal random
variable with $cov(\phi_{s},\phi_{0})=c_{s}$, where $c_{s}$ is defined
as $c_{s}=\frac{1}{\sqrt{n}}\sum_{i=1}^{n}b_{i,s}$. 

Let $\gamma=(\phi_{1}+c_{1}\phi_{0},\cdots,\phi_{n}+c_{n}\phi_{0})$
be a random vector, where $\gamma=MVN(0,\Sigma_{n\times n})$. Letting
$I$ be the $n\times n$ identity matrix and $J$ be the $n\times n$
matrix with all elements equal to $1/n$, we can easily show that
$\Sigma=I-B^{T}JB$ and $\Sigma\Sigma=\Sigma$. Letting $\xi$ be
a random vector, $\xi\sim MVN(0,I_{n\times n})$, we have $\Sigma\xi\sim MVN(0,\Sigma)$
and
\begin{align*}
\sum_{s=1}^{n}\lambda_{s}(\phi_{s}+c_{s}\phi_{0})^{2} & =\gamma^{T}\Lambda\gamma\\
 & =\xi^{T}\Sigma\Lambda\Sigma\xi\\
 & =\xi^{T}B^{T}(B\Sigma B^{T})B\Lambda B^{T}(B\Sigma B^{T})B\xi\\
 & =\xi^{T}B^{T}(I-J)W(I-J)B\xi.
\end{align*}
Because $B\xi\sim MVN(0,I_{n\times n})$, the limiting distribution
of the weighted U is the weighted sum of independent chi-squares,
$U\sim\sum_{s=1}^{n}\lambda_{1,s}\chi_{1,s}^{2}$, where $\{\lambda_{1,s}\}$
are the values of the matrix $(I-J)W(I-J)$.

\section*{Acknowledgements}
This work was supported by the National Institute on Drug Abuse under
Award Number K01DA033346 and by the National Institute of Dental \&
Craniofacial Research under Award Number R03DE022379. The datasets
used for the analyses was obtained from dbGaP through dbGaP accession
number {[}phs000092.v1.p1{]}. 

\section*{Disclosure Declaration}

We declare no conflict of interest.

\begin{supplement}
\sname{Supplementary Material}\label{suppA}
\stitle{Supplementary Material to A Weighted U Statistic for Association Analyses Considering Genetic Heterogeneity}
\slink[url]{http://onlinelibrary.wiley.com/journal/10.1002/(ISSN)1097-0258}
\sdescription{Materials include Supplementary Appendix A to C, Supplementary Simulation I to II, Supplementary Real Data Analysis, Supplementary Table S1 to S9, and Supplementary Figure S1 to S4.}
\end{supplement}

\bibliography{HWU}
\bibliographystyle{apalike} 
\end{document}